\documentstyle[12pt,epsfig]{article}
\newcommand{\be}{\begin{equation}}
\newcommand{\ee}{\end{equation}}
\newcommand{\bea}{\begin{eqnarray}}
\newcommand{\ena}{\end{eqnarray}}
\font\eight=cmr8

\def\pr#1#2#3{ Phys. Rev. ${\bf{#1}}$ (#2) #3}

\def\pl#1#2#3{ Phys. Lett. ${\bf{#1}}$ (#2) #3 }
\def\prep#1#2#3{ Phys. Rep. ${\bf{#1}}$ (#2) #3}
\def\np#1#2#3{ Nucl. Phys. ${\bf{#1}}$ (#2) #3}
\def\zp#1#2#3{ Z. Phys. ${\bf{#1}}$ (#2) #3}

\begin{document}


\begin{center}
{\large \bf Z' PHYSICS}\vspace{0.5cm}\\
Contribution to the workshop on Physics at LEP\,2  
\end{center}
\begin{center}
{\it Conveners}: P.~Chiappetta\footnote{Centre de Physique Th\'eorique, UPR
7061, CNRS-Luminy, Case 907, \\ \null\qquad F-13288 Marseille Cedex 9.} and
C.~Verzegnassi\footnote{Dipartimento di Fisica, Univ. di Lecce, and INFN,
Sezione di Lecce, I-73100 Lecce.}
\end{center}
\begin{center}
{\it Working group}: A. Fiandrino$^1$, J. Layssac\footnote{Physique
Math\'ematique et Th\'eorique, CNRS-URA 768, Univ. de Montpellier II,
\\ \null\qquad F-34095 Montpellier Cedex 5.}, A. Leike\footnote{Sektion Physik,
Ludwig-Maximilians-Univ.M\"unchen, D-80333 M\"unchen, Theresienstr. 37.}, G.
Montagna\footnote{Dipartimento di Fisica Nucleare e Teorica, Univ. di Pavia
and INFN, Sezione di Pavia, \\ \null\qquad I-27100 Pavia.}, O.
Nicrosini\footnote{CERN TH-Division, CH-1211 Gen\`eve 23 (Permanent address :
$^5$).} , F. Piccinini\footnote{INFN, Sezione di Pavia, I-27100 Pavia.}, F.M.
Renard$^3$, S. Riemann\footnote{DESY - Zeuthen, Platanenallee 6,
D-15738 Zeuthen.}, D. Schaile\footnote{Fakult\"at f\"ur Physik,
Albert-Ludwigs-Univ. Freiburg, D-79104 Freiburg.}.
\end{center}
\vspace*{1.0cm}
\tableofcontents
\newpage

\section{Introduction}
Direct production of new particles (with special emphasis on Higgs
and supersymmetric partners) and possible indirect effects due to 
deviations from the predictions of the Standard Model (hereafter denoted as
SM), in particular in the presence of anomalous triple gauge 
couplings, will be soon thoroughly searched for at LEP2 in the 
forthcoming few years. In both cases, typical and sometimes spectacular 
 experimental signatures would exist, allowing to draw unambiguous 
conclusions if a certain type of signal were discovered.
 
	At LEP2, one extra Z (to be called $ Z'$ from now on) would not 
be directly produced, owing to the already existing mass limits from 
Tevatron.Its indirect effects on several observables might be, 
though, sizeable, since it would enter the theoretical expressions 
at tree level.
 In this sense, $ Z'$ effects at LEP2 would be of similar 
type to those coming from anomalous triple gauge couplings
 (hereafter denoted as AGC), although the responsible mechanism 
would be of totally different physical origin.
This peculiar feature of a $ Z'$ at LEP2 has substantially oriented 
the line of research of our working group. In fact, we have tried in 
this report to answer two complementary questions. 

The first one was 
the question "which information on a  $Z'$ can one derive if no indirect
signal of
any type is seen at LEP2? ". To answer this point leads to the derivation, to a
certain  conventionally chosen confidence limit, of (negative) bounds on the 
$Z'$ mass $M_{Z'}$. This has been done for a 
number of "canonical" $Z'$ models, and the resulting bounds (that are 
typically in the TeV range) will be shown in section 2 together with
those for more general $Z'$ 's that might not be detected at an hadronic
collider. 

The second question is: "if a signal of indirect type were seen at LEP2,
 would it be possible to decide whether it may come from a $Z'$ or,
typically,  from a model with anomalous triple gauge couplings?". The
answer to  this question , which is essentially provided from measurements
in the
 final leptonic channel, is given in section 3. In section 4, the role 
of the final WW channel, that might a priori not be negligible for a 
$Z'$ of most general type, has also been investigated and shown to be 
irrelevant at LEP2. Section 5 is devoted to a short final discussion, that 
will conclude our work.

\section{Derivation of bounds.}
Theoretical motivations for the existence of a $Z'$ have already been
given by several authors, and excellent reviews are available 
\cite{review}, where the most studied models are listed and 
summarized. In the following, we will limit ourselves in defining as 
"canonical" cases those of a $Z'$ from  
either $E_6$ \cite{E6} or Left-Right symmetric models \cite{LR} type. 
 For sake of completeness we shall also 
consider the often quoted case of a "Sequential" Standard Model $Z'$ 
\cite{ALT}  
(hereafter denoted as SSM), whose couplings to 
fermions are the same of those of the SM $Z^0$. For these models, 
derivations of bounds for $Z'$ parameters ($Z- Z'$ mixing angle and 
$M_{Z'}$) have been obtained from present data\cite{present},\cite{mix1}
and calculations of discovery limits for $M_{Z'}$  performed for future 
colliders\cite{future},\cite{mix2}, including also a discussion of 
$Z'$ model 
identification. Therefore, the first question that we shall answer in our 
report will be that of how do the LEP2 indirect mass limits compare 
to the direct ones achievable at Tevatron now and in a not too far future 
(i.e. assuming an integrated luminosity of $1 fb^{-1}$). In fact a motivation 
of our work was also that of  deriving limits on a $Z'$ whose couplings to 
fermions are completely  free, including cases that would not be detectable
by any
present or future hadronic collider (for example for negligibly small $Z'$ 
 quark couplings). For this purpose, the final leptonic channel at 
LEP2 provides all the necessary experimental information, and we shall 
consequently begin our analysis with the detailed examination of the 
role of this channel.

To fix our normalization and conventions, we shall write the expression
 of the invariant amplitude for the process $e^+e^- \rightarrow l^+l^-$ 
(where l is a generic charged lepton) in the Born approximation and in
 the presence of a $Z'$. Denoting $q^2$ as the squared center of mass 
energy this amplitude reads in our notations:

\be
A_{ll'}^{(0)}(q^2) = A_{ll'}^{(0) \gamma, Z }(q^2) + A_{ll'}^{(0) Z'}
(q^2)
\ee
where:
\be 
 A_{l{l'}}^{(0) \gamma}(q^2) = i \frac{e_0^2}{q^2} 
{\bar v}_l \gamma_{\mu} u_l {\bar u}_{l'} \gamma^{\mu} v_{l'}
\ee

\be 
 A_{l{l'}}^{(0) Z}(q^2) = \frac{i}{q^2-M^2_Z}  \frac{g_0^2}{4c_0^2} 
{\bar v}_l \gamma_{\mu} ( g_{Vl}^{(0)}-\gamma^5 g_{Al}^{(0)} ) u_l 
{\bar u}_{l'} \gamma_{\mu} (g_{V{l'}}^{(0)}-\gamma^5 g_{A{l'}}^{(0)})
 v_{l'} 
\ee
and(note the particular normalization):
\be 
 A_{ll'}^{(0) Z'}(q^2) = \frac{i}{q^2-M^2_{Z'}}  \frac{g_0^2}{4c_0^2} 
{\bar v}_l \gamma_{\mu} ({g'}_{Vl}^{(0)}-\gamma^5 {g'}_{Al}^{(0)}) u_l 
{\bar u}_{l'} \gamma_{\mu} ({g'}_{Vl'}^{(0)}-\gamma^5 {g'}_{A{l'}}^{(0)}) 
 v_{l'} 
\ee
In the previous equations $e_0 = g_0 s_0$, $c_0^2 = 1-s_0^2$, 
$g_{Al}^{(0)}= I_{3L}=- \frac{1}{2}$ and $g_{Vl}^{(0)}=-
\frac{1}{2}+2s_0^2$. Following the usual approach we shall treat the SM
sector at one loop  and the $Z'$ contribution at tree level. The $Z'$ width
will be assumed  "sufficiently" small 
with respect to $M_{Z'}$ to be safely neglected in the $Z'$ propagator.
 Moreover the $Z- Z'$ mixing angle will be ignored since the limits 
for this quantity provided by LEP1 data from the final leptonic channel
 are enough constraining to rule out the possibility of any observable 
effect in the final leptonic channel at LEP2 (this has been shown in 
a previous paper \cite{mix1} for the "canonical" cases and for a 
general $Z'$ case in a more recent preprint \cite{mix2}).
If we stick ourselves to the final charged leptonic states, we must 
therefore only deal with two "effective" parameters, that might be 
chosen as the adimensional quantities
 $g'_{Vl} \sqrt{\frac{q^2}{M^2_{Z'}-q^2}}$ and 
$g'_{Al} \sqrt{\frac{q^2}{M^2_{Z'}-q^2}}$ (this would be somehow 
reminiscent of notations that are common for models with AGC, with 
$M_{Z'}$ playing the role of the scale 
of new physics). In practice, for the specific purpose of the 
derivation of bounds, a convenient choice was that of the following 
rescaled couplings \cite{rl1}:
\be
v_l^N = g'_{Vl} \sqrt{\frac{q^2}{M^2_{Z'}-q^2}}
 \sqrt{\frac{\alpha}{16c_W^2s_W^4}}
\label{cs5} 
\ee
and 
\be
a_l^N = -g'_{Al} \sqrt{\frac{q^2}{M^2_{Z'}-q^2}}
\sqrt{\frac{\alpha}{16c_W^2s_W^4}}
\label{cs6} 
\ee
with $\alpha = \frac{1}{137}$ and $s_W^2=1- c_W^2=0.231$.

As previously stressed our first task has been that of the derivation 
of constraints for the two previous rescaled couplings from the non 
observation of effects in the leptonic channel. We considered as 
observables at LEP2 the leptonic cross section 
${\sigma}_l$ and the forward-backward asymmetry $A_{FB}^l$, obtained 
from measurements of $\mu$ and  $\tau$ final states, and also the final
$\tau$ polarization $A_\tau$ (we have not yet included the electron channel 
because a full assessment of the corresponding experimental precision
 is more complicated at this stage). Three different energy-integrated 
luminosity configurations were considered, i.e. $ \sqrt{q^2}=140$ GeV 
and $\int L dt=5 pb^{-1}$, 175(500) and 192(300).
Table 1 gives the SM predictions for the three leptonic quantities 
together with the expected experimental accuracies in the three cases. 
 For each energy the first block of three lines contains convoluted 
quantities, whereas the second does not.

A short technical discussion about the way we have calculated 
 the effects of QED emission is now appropriate. In fact two main 
approaches exist that use either complete Feynmann diagrams 
evaluation to compute photonic emission from external legs \cite{bb} or 
the so called QED structure function formalism \cite{sf}, based on 
the analogy with QCD factorization and on the use of the Lipatov-Altarelli-
Parisi evolution equation \cite{russ}. In the calculation of the 
 limits on rescaled parameters performed in this section we have used 
the code ZEFIT \cite{ZEFIT} together with ZFITTER \cite{ZFITTER}.
 These programs utilize the first approach \cite{bb}.
More precisely ZFITTER has all SM corrections and all possibilities to 
apply kinematical cuts.The code ZEFIT contains the additional $Z'$ 
contributions including the full first order QED correction with soft 
photon exponentiation. The first version of this combination, 
applied to LEP1 data,
which  was restricted to definite $Z'$ models, has been adapted for the
model independent analysis that we are now performing at LEP2. 

In practice the 
largest contribution is due to initial state radiation. The 
corresponding expressions for the cross section and forward-backward 
asymmetry read:
\be
\sigma_T = \int^{\Delta}_0 dk \sigma_T^0(s') R_T^e(k)
\ee

\be
A_{FB} = \frac{1}{\sigma_T} \int^{\Delta}_0 dk \sigma_{FB}^0(s') 
R_{FB}^e(k)
\ee
The reduced energy $s'$ reads $s'=q^2(1-k)$ and 
$\Delta=\frac{E_{\gamma}}{E_{beam}}$. 
To first order in $\alpha$, improved by soft photon exponentiation, 
the two functions  $R_T^e(k)$ and $R_{FB}^e(k)$ are given by the 
following expressions:
\be
R_{T,FB}^e(k) = (1+S_e) \beta_e k^{\beta_e -1}+ H_{T,FB}^e (k)
\ee
where
\be
 \beta_e = 2 \frac{\alpha}{\pi} e_e^2  \ln( \frac{q^2}{m_e^2}-1)\ ,
\ee
the soft radiaton part reads:
\be
 S_e = \frac{\alpha}{\pi} e_e^2 \left( \frac{\pi^2}{3}-\frac{1}{2}+
\frac{3}{2} \ln( \frac{q^2}{m_e^2}-1) \right)
\ee
and the hard radiation parts:
\be
 H_T(k) = \frac{\alpha}{\pi} e_e^2 \left\{ \frac{1+(1-k)^2}{k}
 \ln( \frac{q^2}{m_e^2}-1) \right\} -\frac{\beta_e}{k}
\ee

\be
 H_{FB}(k) = \frac{\alpha}{\pi} e_e^2 \left( \frac{1+(1-k)^2}{k}
\frac{1-k}{(1-\frac{k}{2})^2}
 ( \ln \frac{q^2}{m_e^2}-1-\ln \frac{1-k}{(1-\frac{k}{2})^2}) \right) 
-\frac{\beta_e}{k}
\ee
The value of $\Delta$ is chosen by requiring that the invariant mass 
of the final fermion pair $M_{ll}^2=(1-\Delta) q^2$ is "sufficiently"
 greater than $M_Z$, to exclude the radiative return to the $Z$ peak.
 This has very important implications for searches of $Z'$ effects, 
since it is well known that the radiative tail can enhance the SM cross 
section by a factor $2-3$, then completely diluting the small $Z'$ 
effects, as fully discussed in a previous paper \cite{djouadi} . 
Results shown in table 1 are obtained for an  invariant 
mass of the final fermion pair $M_{ll}$ greater than $120$ GeV. 

One clearly sees from inspection of Table 1 that the most promising of
the three energy-luminosity combinations for what concerns the relative size
of the error is $175$ GeV and $500 pb^{-1}$.

>From now on  we shall therefore  
concentrate on this configuration and evaluate the bounds on $Z'$ 
rescaled parameters that would follow from the non observation of any 
virtual effect. 
With this purpose we have made full use of the code ZEFIT and chosen 
$\Delta=.7$ ,although smaller values like for instance the one used 
in table 1 would lead to practically the same conclusions.
 
To obtain exclusion limits, we calculate the SM predictions of all 
observables $O_i(SM)$ and compare them with the prediction 
 $O_i(SM,v_i^N,a_i^N)$ from a theory including a $Z'$. In our fits we 
use the errors $\Delta O_i$ calculated using the same assumptions as 
in Table 1 and define: 

\be
\chi^2 =  \sum_{O_i} \left( \frac{O_i(SM)-O_i(SM,Z')}{\Delta O_i}
 \right)^2.
\label{chi2}
\ee
${\chi}^2 < \chi^2_{min} + 5.99$ corresponds to $95\%$ confidence 
level for one sided exclusion bounds for two parameters.

\newpage

\null
\vskip 3cm


\hskip -8mm\begin{tabular}{|r|r|r|r|r|r|r|r|r|r|r|r|r|}
\hline
&$E_{cm}$&Lumi&Acc&$\sigma(pb)$&$\Delta\sigma_{stat}$&$\Delta
\sigma_{syst}$&$\Delta\sigma$&Error&$A_{FB}$&$(\Delta
A_{FB})$&$A_{\tau}$&$(\Delta A_{\tau})$\\
\hline
$\mu$&140.&5.&.90&6.43&1.20&.05&1.20&18.6\%&.684&.136&&\\
$\ell$&140.&5.&.90&6.51&0.85&.04&0.85&13.1\%&.684&.095&&\\
$\tau$&140.&5.&&&&&&&&&-.104&0.61\\
\hline
$\mu$&140.&5.&.90&6.98&1.25&.05&1.25&17.9\%&.666&.133&&\\
$\ell$&140.&5.&.90&6.98&0.88&.04&0.88&12.6\%&.666&.094&&\\
$\tau$&140.&5.&&&&&&&&&-.102&0.59\\
\hline
$\mu$&175.&500.&.90&4.13&0.10&.03&0.10&2.4\%&.602&.019&&\\
$\ell$&175.&500.&.90&4.15&0.07&.03&0.07&1.7\%&.602&.013&&\\
$\tau$&175.&500.&&&&&&&&&-.082&0.08\\
\hline
$\mu$&175.&500.&.90&4.01&0.09&.03&0.10&2.5\%&.586&.01&&\\
$\ell$&175.&500.&.90&4.01&0.07&.02&0.07&1.8\%&.586&.01&&\\
$\tau$&175.&500.&&&&&&&&&-.079&0.08\\
\hline
$\mu$&192.&300.&.90&3.47&0.11&.02&0.12&3.3\%&.579&.02&&\\
$\ell$&192.&300.&.90&3.49&0.08&.02&0.08&2.4\%&.579&.01&&\\
$\tau$&192.&300.&&&&&&&&&-.085&0.11\\
\hline
$\mu$&192.&300.&.90&3.28&0.11&.02&0.11&3.4\%&.565&.02&&\\
$\ell$&192.&300.&.90&3.28&0.08&.02&0.08&2.5\%&.565&.02&&\\
$\tau$&192.&300.&&&&&&&&&-.081&0.11\\
\hline
\end{tabular}
\begin{table}[hbtp]
\caption{SM predictions for leptonic observables including
experimental  accuracies.
As a simple simulation of the detector acceptance, 
we impose that the angle between the outgoing leptons and the beam 
axis is larger than $20^{\circ}$, leading to an acceptance of about $0.9$. 
The first line gives the muon cross section and the forward-backward 
asymmetry and errors. The second line 
gives the averaged $\mu$ and $\tau$ cross section and asymmetry (and 
errors) whereas the third line contains the $\tau$ polarization(obtained 
by using only the $\rho$ and  $\pi$ channels and assuming an average 
sensitivity of $.5$). Concerning systematics we assumed $.5\%$ relative 
error for $\mu$ and $\tau$ selections and also for luminosity. For all 
asymmetries we did not consider any systematic error. 
All quoted errors refer to a single LEP experiment. Taking into account 
the type of systematic errors and the relative size of systematics vs 
statistics, it is a good approximation to divide the error by 2 to 
estimate the combined error of the four experiments.}
\end{table}


\newpage

\begin{figure}

\hbox to\hsize{\epsfxsize=0.5\hsize
\epsffile[0 0 340 340]{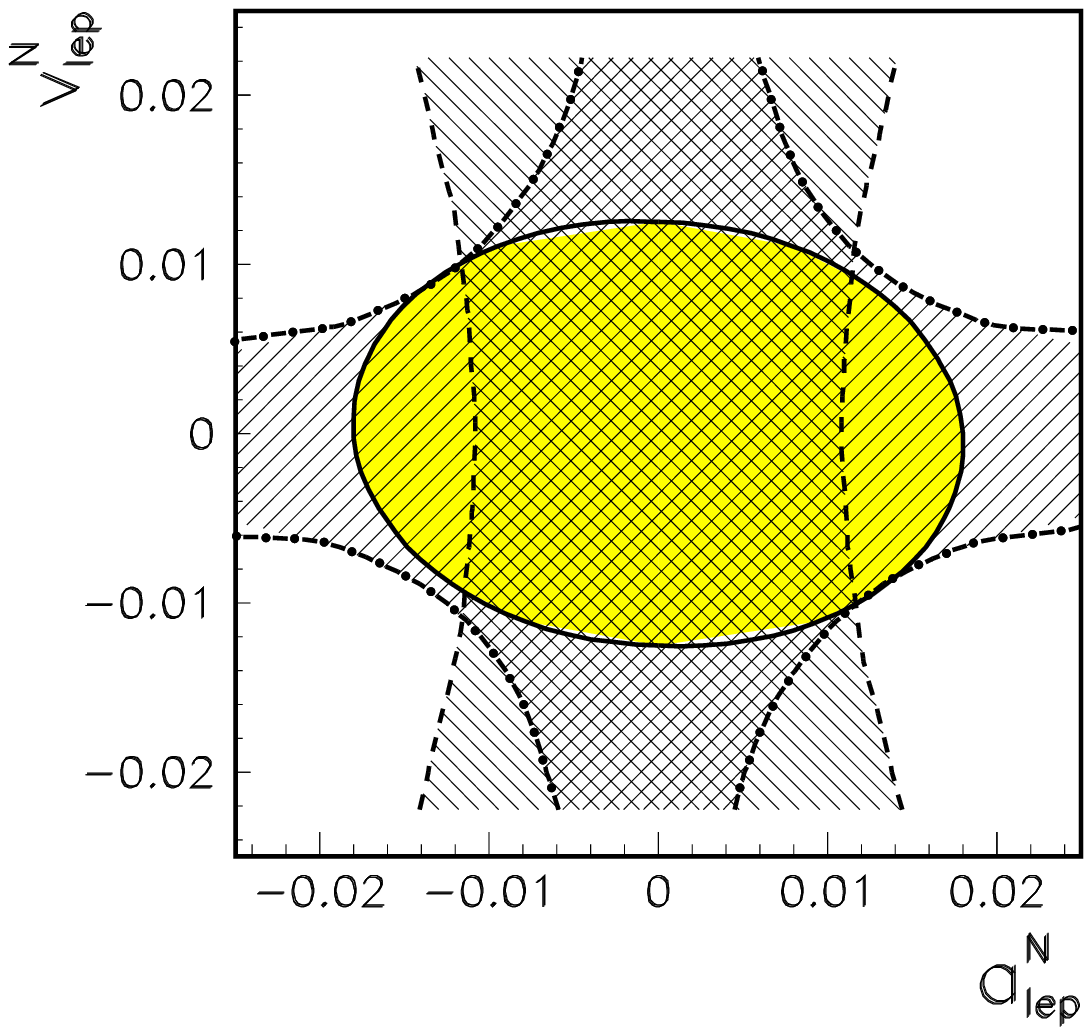}
\hfill\epsfxsize=0.5\hsize
\epsffile[0 0 340 340]{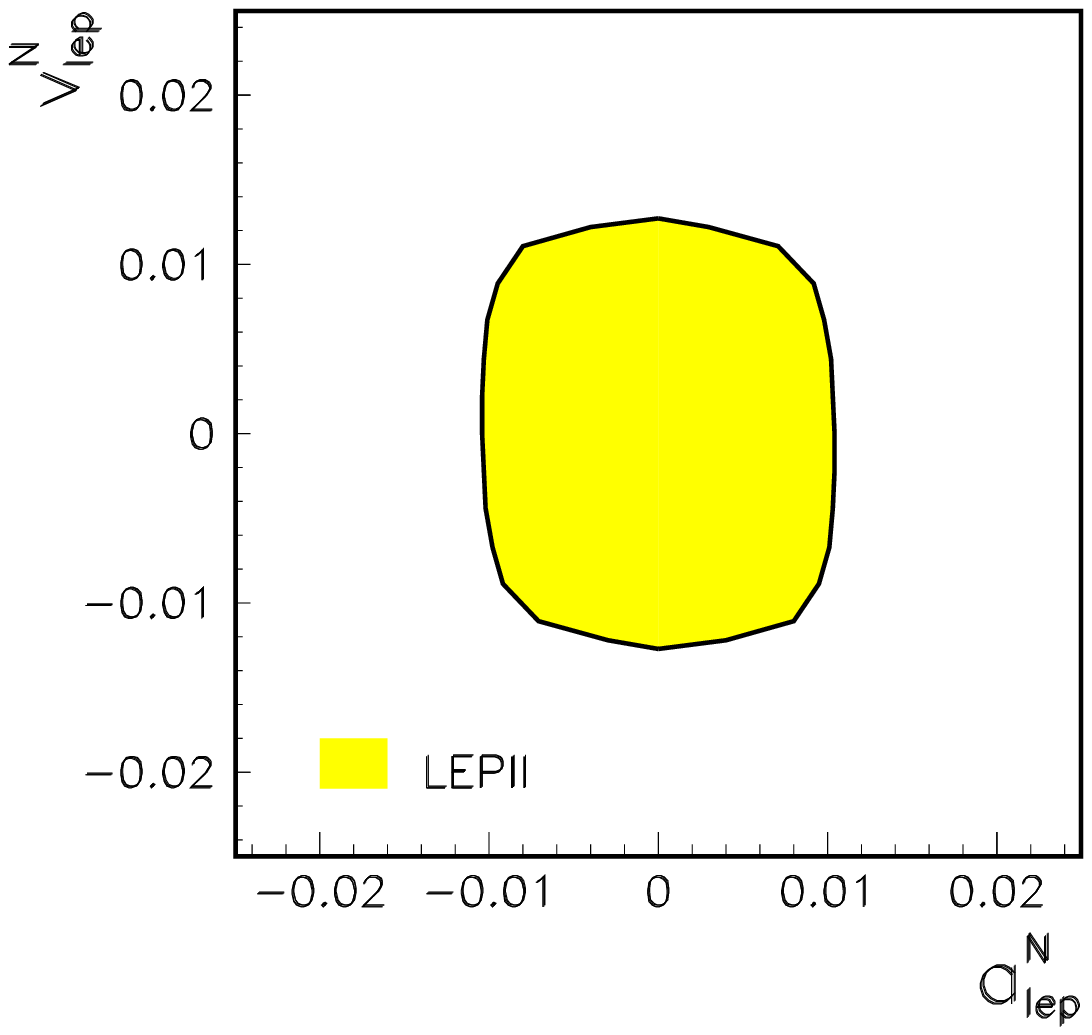}}

\noindent {\bf Fig.~1} {\eight The areas in the $(a^N_{\ell},v^N_{\ell})$
plane excluded with
$95\%$ CL at LEP 2 by different observables i.e. $\sigma_{\ell}$
(ellipse) and $A^l_{FB}$ (crossed lines). The remaining contours, that do
not improve the limits, would correspond to a measurement of $A_{\tau}$
with an accuracy twice better than the realistic one quoted in Table~1.\\
\noindent{\bf Fig.~2} The area in the $(a^N_l,v^N_l)$ plane excluded
with
$95\%$ confidence at LEP 2 by the combination of all the leptonic
observables.}
\end{figure}

 Figures 1 and 2 give our model independent constraints to
 the rescaled leptonic $Z'$ couplings including all radiative corrections
, i.e. the QED radiation and the electroweak corrections, that have 
a very small influence on the results. In figure 1 the constraint 
from each observable is shown separately. The combined
exclusion region is depicted in the  next figure 2 and 
a few comments on the previous figure are now appropriate. It 
represents in fact the most general type of constraints that can be 
derived on a $Z'$ from the absence of signals in the leptonic channel 
at LEP2, under the assumption that the $Z'$ couples to charged leptons 
in a universal way. In particular, from this figure one 
might derive bounds on the parameters of  $Z'$ that were only coupled 
to leptons  and would therefore escape detection at any hadronic machine. 
For such models, the limit on $M_{Z'}$ would be then derivable to a 
very good(and conservative) approximation, for given $Z'$ couplings, by 
the simple expression(derived from  eq.~(\ref{cs5}) and eq.~(\ref{cs6})):
\be
\frac{M_{Z'}^2}{q^2}\geq \frac{1}{r^2}
\frac{{g'}_{Vl}^2+{g'}_{Al}^2}{\frac{\alpha}{16c_w^2s_w^4}}
\ee
where r is the distance from the origin in the ($v_l^N,a_l^N$) plane of 
the intersection between the boundary region of figure 2 and the 
straight line:  
 \be
 \frac{g'_{Vl}}{g'_{Al}} = -  \frac{v_l^N}{a_l^N} = k
\ee
where k is fixed by the considered model and r will vary between $.01$ and 
$.015$.
(Numerically $\frac{\alpha}{16c_w^2s_w^4} \sim 81$.)


\vskip 15mm

\hskip -10mm\begin{tabular}{|r|r|r|r|r|r|r|r|r|r|r|r|r|}
\hline
&$E_{cm}$&Lumi&Acc&$\sigma(pb)$&$\Delta\sigma_{stat}$&$\Delta
\sigma_{syst}$&$\Delta\sigma$&Error&$A_{FB}$&$(\Delta
A_{FB})$&$R$&$Error$\\
\hline
$b$&140.&5.&0.25&10.46&2.89&.31&2.91&27.8\%&.499&.379&&\\
$h$&140.&5.&1.00&59.03&3.44&.66&3.50&5.9\%&&&&\\
$R_h$&140.&5.&&&&&&&&&9.181&14.3\%\\
$R_b$&140.&5.&&&&&&&&&0.177&28.4\%\\
\hline
$b$&140.&5.&0.25&10.62&2.92&.32&2.93&27.6\%&.509&.373&&\\
$h$&140.&5.&1.00&60.49&3.48&.68&3.54&5.9\%&&&&\\
$R_h$&140.&5.&&&&&&&&&8.666&13.9\%\\
$R_b$&140.&5.&&&&&&&&&0.176&28.2\%\\
\hline
$b$&175.&500.&0.25&5.15&0.20&.15&.26&5.0\%&.543&.052&&\\
$h$&175.&500.&1.00&31.24&0.25&.35&.43&1.4\%&&&&\\
$R_h$&175.&500.&&&&&&&&&7.572&2.1\%\\
$R_b$&175.&500.&&&&&&&&&0.165&5.1\%\\
\hline
$b$&175.&500.&.25&4.69&0.19&.14&0.24&5.1\%&.562&.054&&\\
$h$&175.&500.&1.00&28.90&0.24&.32&0.40&1.4\%&&&&\\
$R_h$&175.&500.&&&&&&&&&7.215&2.1\%\\
$R_b$&175.&500.&&&&&&&&&0.162&5.3\%\\
\hline
$b$&192.&300.&0.25&4.08&0.23&.12&.26&6.5\%&.554&.079&&\\
$h$&192.&300.&1.00&25.22&0.29&.28&.40&1.6\%&&&&\\
$R_h$&192.&300.&&&&&&&&&7.265&2.8\%\\
$R_b$&192.&300.&&&&&&&&&0.162&6.6\%\\
\hline
$b$&192.&300.&.25&3.62&0.22&.11&0.25&6.8\%&.577&.078&&\\
$h$&192.&300.&1.00&22.79&0.28&.25&0.38&1.6\%&&&&\\
$R_h$&192.&300.&&&&&&&&&6.942&2.9\%\\
$R_b$&192.&300.&&&&&&&&&0.159&6.9\%\\
\hline
\end{tabular}
\begin{table}[hbtp]
\caption{SM predictions for hadronic observables including
experimental  accuracies.
The first line gives the b quark cross section and forward-backward 
asymmetry and the corresponding experimental errors. The second line 
gives the total hadronic cross section (and errors) whereas the 
third line contains the ratio $R_h=\frac{\sigma_h}{\sigma_l}$ 
and the fourth line the ratio $R_b=\frac{\sigma_b}{\sigma_h}$
. Concerning systematics we assumed $1\%$ relative 
error for hadron selection and $3\%$ relative error for
 b quark selection.We did not consider
 any systematic error for all asymmetries. Concerning the b quark cross
section we assume a tagging efficiency of $25\%$ (vertex tag) and 
 $10\%$ (lepton tag) for the asymmetry. The first block of 
numbers (upper four lines) refer to convoluted quantities where proper
cuts have been applied, whereas the lower four lines contain deconvoluted 
quantities.}
\end{table}

\newpage

In a less special situation, the $Z'$ couplings to quarks will 
not be vanishing. In these cases, to derive meaningful bounds, the 
full information coming from the final hadronic channel should be also  
exploited. At LEP2, we assumed the availability of three different 
measurements, i.e. those of the total hadronic cross section 
${\sigma}_h$ and those of the cross section and forward-backward 
asymmetry for $b \bar b$ production,${\sigma}_b$ and $A_{FB}^b$. In 
table 2 we give the related expected experimental accuracies, for the 
three energy-luminosity configurations already investigated for the 
final leptonic channel in Table 1, and under the same general 
assumptions listed in the discussion preceding the presentation of 
this table.  

>From the combination of the leptonic and hadronic channels, a fully 
general investigation of the six rescaled $Z'$ couplings (there would 
be four extra rescaled couplings for "up" and "down" type quarks) might be,
 in principle, carried through if at least four hadronic independent 
observables were measured at LEP2. This could be obtained if one more 
hadronic asymmetry were measured. In practice, though, the utility of 
such an approach is somehow obscured by practical considerations( like 
the realistic achievable experimental accuracy). For these reasons, we 
have therefore decided to make full use of the hadronic observables to 
derive limits on $M_{Z'}$ only for a number of "canonical" models 
where the $Z'$ couplings to fermions are constrained.
As relevant examples to be investigated, we shall discuss $E_6$ models 
\cite{E6} and Left-Right symmetric models \cite{LR}, for which the 
$Z'$ current can be decomposed as:
 \be
 J_{Z'}^{\mu} = J_{\chi}^{\mu} \cos \beta +  J_{\psi}^{\mu} \sin \beta 
\ee
 or
 \be
 J_{Z'}^{\mu} = J_{LR}^{\mu} \alpha_{LR} - J_{B-L}^{\mu} \frac{1}
{2 \alpha_{LR}} 
\ee

\begin{table}[hbtp]
\begin{center}

Table 3a

\medskip

\begin{tabular}{|c|c|c|}
\hline
f&$\frac{{g'}_{Vf}}{s_w}$&$\frac{{g'}_{Af}}{s_w}$\\
\hline
 l &$\frac{2}{\sqrt 6} \cos \beta$&$\frac{1}{\sqrt 6} \cos \beta + 
\frac{\sqrt 10}{6} \sin \beta $\\
\hline
 u &0&$-\frac{1}{\sqrt 6} \cos \beta + \frac{\sqrt 10}{6} \sin \beta $\\
\hline
 d &$-\frac{2}{\sqrt 6} \cos \beta$&$\frac{1}{\sqrt 6} \cos \beta + 
\frac{\sqrt 10}{6} \sin \beta $\\
\hline
\end{tabular}

\vskip 4truemm

Table 3b

\medskip

\begin{tabular}{|c|c|c|}
\hline
f&$\frac{{g'}_{Vf}}{s_w}$&$\frac{{g'}_{Af}}{s_w}$\\
\hline
 l &$\frac{1}{\alpha_{LR}}- \frac{\alpha_{LR}}{2}$&$\frac{\alpha_{LR}}{2}$\\ 
\hline
 u &$-\frac{1}{3 \alpha_{LR}} +
\frac{\alpha_{LR}}{2}$&$-\frac{\alpha_{LR}}{2}$\\ 
\hline
 d &$-\frac{1}{3 \alpha_{LR}} - \frac{\alpha_{LR}}{2}$&$\frac{\alpha_{LR}}{2}$\\
\hline
\end{tabular}

\caption{Couplings of ordinary fermions (f=l,u,d) to $Z'$ boson 
a) from $E_6$ models as a function of the parameter $\cos \beta$
b) from Left-Right models as a function of the parameter $\alpha_{LR}$.}
\end{center}
\end{table}


\vspace{0.3cm}

In table 3 we have given the $Z'$ couplings to l, u and d fermions for
the two models. Some specific relevant cases in the $E_6$ sector are 
the so called $\chi$ model (corresponding to $\cos \beta =1$), $\psi$ model
($\cos \beta =0$) and  $\eta$ model
($\arctan \beta = - \sqrt {\frac{5}{3}}$). Special 
cases for Left-Right symmetric models are obtained for
 $\alpha_{LR} = \sqrt {\frac{2}{3}}$ (this case reproduces the $\chi$ 
 model) and  $\alpha_{LR} = \sqrt 2$ (the so called manifestly L-R 
symmetric model). Finally we also chose the $Z'$ of the Sequential 
Standard Model(which has the same fermionic couplings as those of the 
SM Z) as an additional benchmark.

Table 4 shows the  CL bounds on  $M_{Z'}$ obtainable from the 
non observation of any effect at LEP2 in the configuration: $175$ GeV, 
 $500 pb^{-1}$. In fact one can easily show that for virtual 
$Z'$ searches this configuration is the best of the three that 
we have considered for LEP2, since the simple scaling law for the achievable
 limit ${(M_{Z'})}_{max} \sim {\left( q^2 \int L \right)}^{\frac{1}{4}}$ 
 applies. The different lines show the influence of the 
hadronic observables. As one sees, this is indeed relevant for the 
SSM $Z'$. In the other cases it improves the bounds derived from purely
 leptonic observables by an amount of less than (typically) a relative 
$10\%$.

\begin{table}[hbtp]
\begin{center}
\begin{tabular}{|l|ll|ll|ll|ll|l|}
\hline
&$\chi$&&$\psi$&&$\eta$&&LR&&SSM\\
\hline
$\sigma_{\ell},A^{\ell}_{FB}$&870&&640&&525&&838&&1238\\
\hline
$\sigma_{\ell},A^{\ell}_{FB},\sigma_{had}$&900&&642&&550&&854&&1530\\
\hline
$\sigma_{\ell},A^{\ell}_{FB},R_b,A^{b}_{FB},\sigma_{had}$&930&&666&&560&&880
&&1580\\
\hline
\end{tabular}

\caption{Maximal $Z'$ masses $M_Z'$ excluded by leptonic and hadronic
observables.
${\chi}^2 < \chi^2_{min} + 2.7$ ($95\%$ CL, one sided limits).}
\end{center}
\end{table}

The more general analysis of the two models of extra gauge, that 
corresponds to values of $\cos \beta$ ranging from $-1$ to $+1$ 
(positive  $\sin \beta$) and   $\alpha_{LR}$ ranging from 
$\sqrt {\frac{2}{3}}$ to $\sqrt 2$, has been summarized in figures 3 
and 4 (full lines). One sees that the best $ M_{Z'}$ limits correspond 
to models where  $\cos \beta \sim1$ and where $\alpha_{LR}$ is at the 
boundary of its allowed interval, for which the bounds derivable at 
LEP2 would be about 1 TeV.

\newpage

\begin{figure}

\hbox to\hsize{\epsfxsize=0.5\hsize
\epsffile[8.5 22.0 541.3 747.1]{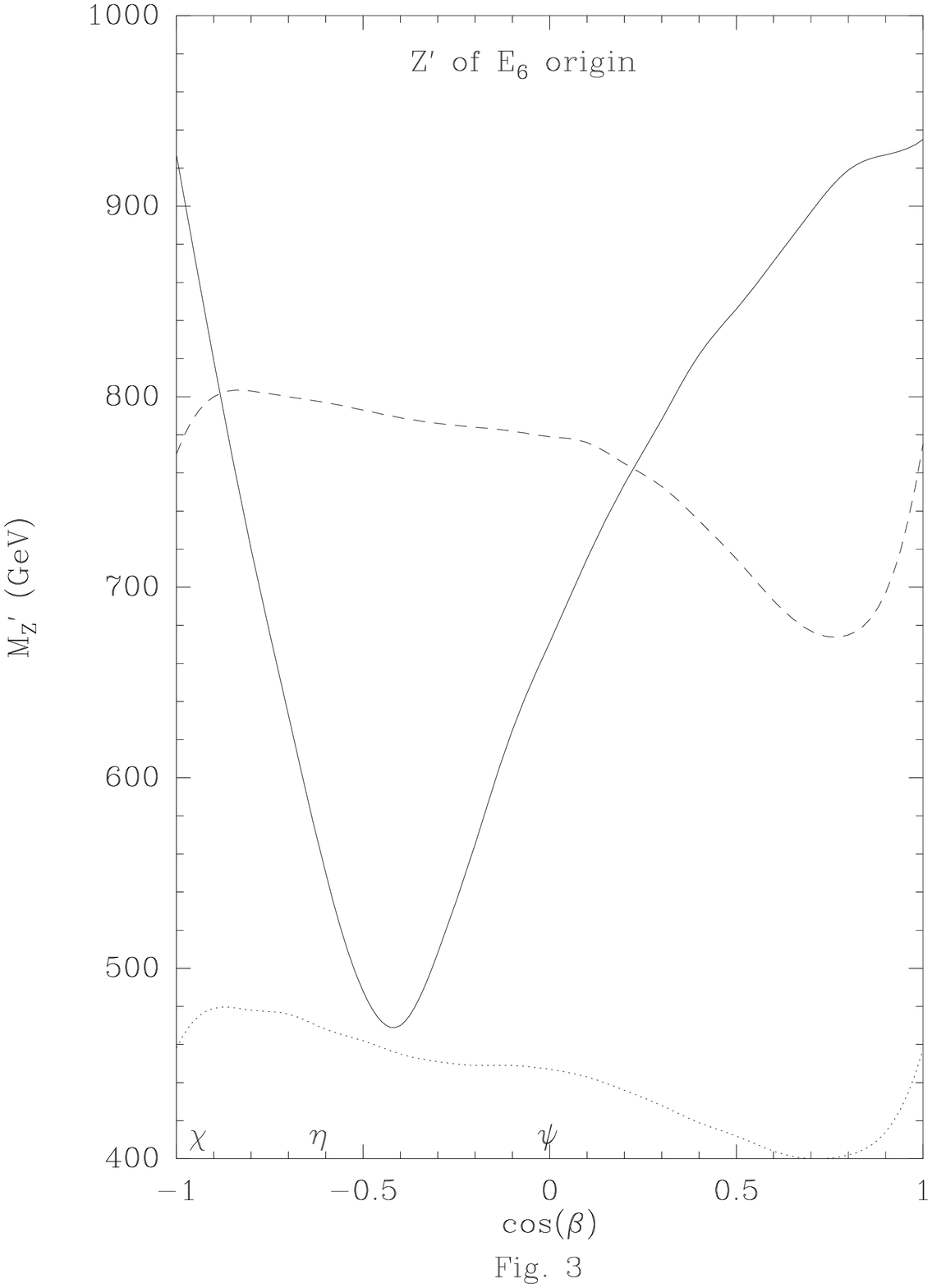}
\hfill\epsfxsize=0.5\hsize
\epsffile[8.5 22.0 541.3 747.1]{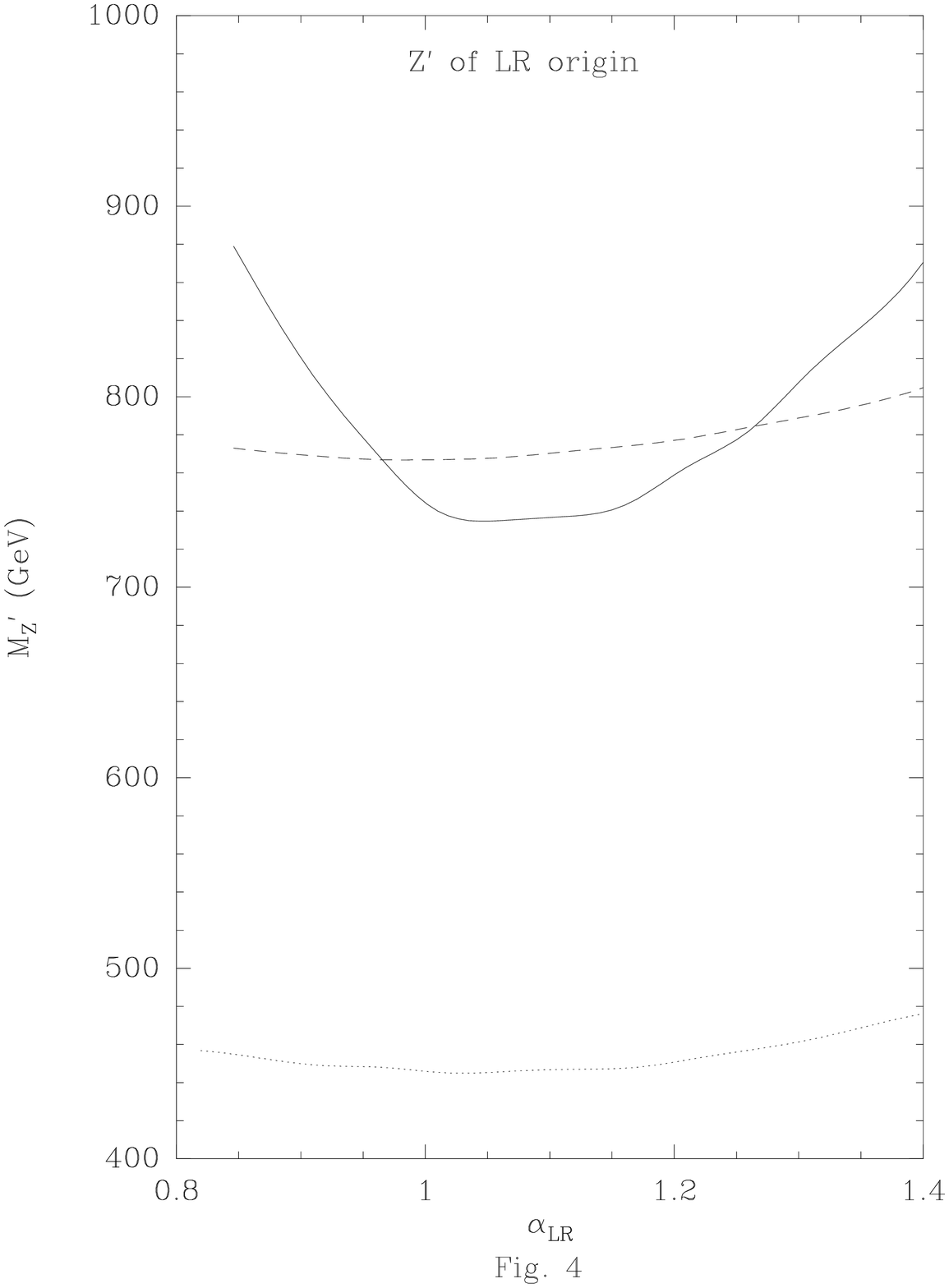}}

\noindent {\eight Maximal $Z'$ masses excluded at LEP2 (full
curve) and at  Tevatron with a luminosity of $1 fb^{-1}$ (dashed curve) or $20
pb^{-1}$ (dotted curve) for $E_6$ models (Fig.~3) and for Left-Right models
(Fig.~4).}

\end{figure}

The values that we have derived should be compared with those 
already available and with those reachable 
 in a not too far future at Tevatron. To fix the scales for the 
comparison, we have considered the limits that would correspond to an 
energy of $1.8$ TeV with an integrated luminosity of  $1 fb^{-1}$ and 
drawn on the same figures 3 and 4 (dashed lines) the expected Tevatron 
limits, that would "compete" with the LEP2 results (the present Tevatron 
limits for $20 pb^{-1}$  correspond to the dotted curves). The  
limits correspond to $95 \%$ CL bounds on $M_{Z'}$ based on 10 events 
in the $e^+e^- + \mu^+ \mu^-$ channel, assuming that $Z'$ can only 
decay in the three conventional fermion families. The values  that we 
plot are in agreement with those quoted in a recent report  \cite{CVETIC}. 
One sees from figures 3 and 4 that for the $E_6$ models the LEP2 
limits are in a sense complementary to those of Tevatron in the 
future configuration, providing better or worse indications depending 
on which range is chosen for $\cos \beta$. LEP2 is better if 
$\cos \beta$ lies in the vicinity of $-1$ and $\cos \beta \geq 0.4$. 
On the contrary, for Left-Right symmetric models LEP2 appears to do 
much better, except for $\alpha_{LR}$ ranging between $1$ and $1.2$ 
where Tevatron could provide limits a bit higher;
concerning the SSM $Z'$, LEP2 does systematically better since 
it can reach $1.5$ TeV whereas the future Tevatron limit 
is around $900$ GeV. Note that, should other exotic or supersymmetric
channel be open for $Z'$ decay, the Tevatron limits might decrease by 
as much as $30 \%$, depending on the considered model \cite{CVETIC}. 
We conclude therefore that, until the Tevatron luminosity will reach 
values around $10 fb^{-1}$, the canonical LEP2 bounds will be, least 
to say, strongly competitive.
 
Our determination of bounds is at this point finished for what concerns 
the final fermionic channel. In the next section we shall try to derive 
some model-independent criterion to identify $Z'$ signals at LEP2.
  
\section{Search for signals: the leptonic channel.}
In this section, we shall assume that some virtual signal has been seen 
at LEP2 in the leptonic channel. In this case, we shall show that it would 
be possible to conclude whether this signal is due to a $Z'$ or not.

 This can be easily understood if one 
compares the $Z'$ effect to a description that includes the SM 
effects at one loop, and we shall briefly summarize the main points.
For what concerns the treatment of the SM sector, a prescription has 
been very recently given \cite{CV2}, that corresponds to a 
"Z-peak substracted" representation of four fermion processes, in 
which a modified Born approximation and "substracted" one loop 
corrections are used. These corrections, that are "generalized" 
self-energies, i.e. gauge-invariant combinations of self-energies, 
vertices and boxes, have been called in \cite{CV2} (to which we refer 
for notations and conventions) 
$\widetilde {\Delta} \alpha (q^2)$, $R(q^2)$ and $V(q^2)$ respectively.
 As shown in ref \cite{CV2}, 
they turn out to be particularly useful whenever effects of new physics 
must be calculated. In particular, the effect of a general $Z'$ would  
appear in this approach as a particular modification of purely "box" 
type to the SM values of $\widetilde {\Delta} \alpha (q^2)$, $R(q^2)$ 
and $V(q^2)$ given by the following prescriptions:
\be
{\widetilde {\Delta}}{\alpha} ^{(Z')} (q^2) = - \frac{q^2}{M^2_{Z'}-q^2}
\frac{1}{4c_1^2s_1^2} g^2_{Vl} {(\xi_{Vl} - \xi_{Al})}^2
\label{rese1} 
\ee

\be
R^{(Z')}(q^2) = (\frac{q^2-M^2_Z}{M^2_{Z'}-q^2}) \xi_{Al}^2
\label{rese2}
\ee

\be
V^{(Z')}(q^2) = -(\frac{q^2-M^2_Z}{M^2_{Z'}-q^2}) 
\frac{g_{Vl}}{2c_1s_1} \xi_{Al} (\xi_{Vl} - \xi_{Al})
\label{rese3}
\ee
where we have used the definitions:
\be
 \xi_{Vl} = \frac{g'_{Vl}}{g_{Vl}}
\label{res3}
\ee

\be
 \xi_{Al} = \frac{g'_{Al}}{g_{Al}}
\label{res4}
\ee

with $g_{Vl}= \frac{1}{2} (1-4s_1^2)$;$g_{Al}= -\frac{1}{2}$ and 
$c_1^2s_1^2 = \frac{\pi \alpha (0)}{\sqrt 2 G_{\mu} M^2_Z}$. 
To understand the philosophy of our approach it is convenient to write 
the expressions at one-loop of the three independent leptonic 
observables that will be measured at LEP2, i.e. the leptonic cross section, 
the forward-backward asymmetry and the final $\tau$ polarization. 
Leaving aside specific QED corrections extensively discussed in the previous
section, these  expressions read:

\bea  \sigma_l(q^2)=&&\sigma^{Born}_l(q^2)\bigm\{1+{2\over
\kappa^2(q^2-M^2_Z)^2+q^4}[\kappa^2(q^2-M^2_Z)^2
\tilde{\Delta}\alpha(q^2)\nonumber\\
&&-q^4(R(q^2)+{1\over2}V(q^2))]\bigm\}
\label{rese4} \ena

\bea  A^l_{FB}(q^2)=&&A^{l,Born}_{FB}(q^2)\bigm\{1+
{q^4-\kappa^2(q^2-M^2_Z)^2
\over\kappa^2(q^2-M^2_Z)^2+q^4}[
\tilde{\Delta}\alpha(q^2)+R(q^2)]\nonumber\\
&&+{q^4\over\kappa^2(q^2-M^2_Z)^2+q^4}V(q^2)]\bigm\}
\label{rese5} \ena

\bea A_{\tau}(q^2)=&&A^{Born}_{\tau}(q^2)
\bigm\{ 1+[{\kappa(q^2-M^2_Z)
\over\kappa(q^2-M^2_Z)+q^2}-{2\kappa^2(q^2-M^2_Z)^2\over
\kappa^2(q^2-M^2_Z)^2+q^4}]
[\tilde{\Delta}\alpha(q^2)\nonumber \\
&&+R(q^2)]
-{4c_1s_1\over v_1}V(q^2) \bigm\}
\label{rese6} \ena

where $\kappa$ is a numerical constant 
($\kappa^2 =  {(\frac{\alpha}{3 \Gamma_l M_Z})}^2 \simeq 7$) and we 
refer to \cite{CV2} for a more detailed derivation of the previous 
formulae.
 
A comparison of  eqs.~(\ref{rese4}-\ref{rese6}) with eqs.~(\ref{rese1}
-\ref{rese3}) shows that, in the three leptonic observables, only two 
effective parameters, that could be taken for instance as 
$\xi_{Vl} \frac{M_Z} {\sqrt{M^2_{Z'}-q^2}}$ and 
$(\xi_{Vl} - \xi_{Al})\frac{M_Z} {\sqrt{M^2_{Z'}-q^2}}$(to have 
dimensionless quantities, other similar definitions would do equally 
 well), enter. This leads to the conclusion that it must be possible 
to find a relationship between the relative $Z'$ shifts 
$\frac{\delta \sigma_l}{\sigma_l}$, $\frac{\delta A^l_{FB}}{A^l_{FB}}$ 
 and  $\frac{\delta A_{\tau}}{A_{\tau}}$ (defining, for each observable
 $O_i =O_i^{SM} + \delta O_i^{Z'}$) that is completely independent of 
the values of these effective parameters. This will correspond to a 
region in the 3-d space of the previous shifts that will be fully 
characteristic of a model with the most general type of $Z'$ that we 
have considered. We shall call this region "Z' reservation".

To draw this reservation would be rather easy if one relied on a 
calculation in which the $Z'$ effects are treated in first 
approximation, i.e. only retaining the leading effects, and not taking 
into account the QED radiation. After a rather straightforward 
calculation one would then be led to the following approximate 
expressions that we only give for indicative purposes:
\be
({\frac{\delta A_{\tau}}{A_{\tau}}})^2 \simeq f_1 f_3
\frac{8 c_1^2 s_1^2}{v_1^2} \frac{\delta \sigma_l}{\sigma_l} 
\left( \frac{\delta A^l_{FB}}{A^l_{FB}} +
\frac{1}{2} f_2 \frac{\delta \sigma_l}{\sigma_l} \right)
\label{rese7}  
\ee

where the $f_i$'s are numerical constants, whose expressions can be found
in~\cite{CV2}.

Eq.~(\ref{rese7}) is an approximate one. A more realistic description 
can only be obtained if the potentially dangerous QED effects are 
fully accounted for.In order to accomplish this task, the QED structure 
function formalism\cite{sf} has been employed as a reliable tool for 
the treatment of large undetected initial-state photonic radiation. 
Using the structure function method amounts to writing, in analogy with 
QCD factorization, the QED corrected cross section as a convolution 
of the form:

\be
\sigma (q^2) = \int dx_1 dx_2 D(x_1,q^2) D(x_2,q^2) 
\sigma_0 ((1-x_1x_2)q^2) (1+ {\delta}_{fs}) \Theta (cuts)
\label{rese8}
\ee
where $\sigma_0$ is the lowest order kernel cross section, taken at 
the energy scale reduced by photon emission,$ D(x,q^2)$ is the electron
(positron) structure function, ${\delta}_{fs}$ is the correction factor 
taking care of QED final-state radiation and $\Theta (cuts)$ represents 
the rejection algorithm to implement possible experimental cuts. Its 
expression, obtained by solving the Lipatov-Altarelli-Parisi evolution equation
 in the non-singlet approximation, can be found in \cite{russ} together 
with a complete discussion of the method. 
In order to proceed with the numerical simulation of the $Z'$ effects
under realistic experimental conditions, the master formula 
eq.~(\ref{rese8}) has  been implemented in a Monte Carlo event 
generator which has been first checked against currently used LEP1 
software \cite{nicrosini}, found to be in very good agreement and then 
used to produce our numerical results. The $Z'$ contribution has been 
included in the kernel cross section $\sigma_0$ computing now the 
s-channel Feynman diagrams associated to the production of a leptonic 
pair in $e^+e^-$ annihilation mediated by the exchange of a photon, a SM 
Z and an additional $Z'$ boson. In the calculation, which has been 
carried out within the helicity amplitude formalism for massless fermions and 
 with the help of the program for algebraic manipulations SCHOONSCHIP 
\cite{veltman}, the $Z'$ propagator has been included in the 
zero-width approximation. Moreover, the bulk of non QED corrections has 
been included in the form of the Improved Born Approximation, choosing 
 $\bar \alpha (s), M_Z, G_F$ and $\Gamma_Z$ as input parameters. The 
values used for the numerical simulation are \cite{SMLEP}: 
$M_Z= 91.1887$ GeV, $\Gamma_Z =2.4979$ GeV. The center of mass energy 
has been fixed to  $\sqrt q^2=175$ GeV and the cut $x_1x_2 > 0.35$ (that 
would correspond to the choice $\Delta = 0.65$ in the notations of the 
previous section) has been imposed in order to remove the events due to 
Z radiative return and hence disentangle the interesting virtual $Z'$ 
effects. These have been investigated allowing the previously defined 
ratios $\xi_{Vl}$ and $\xi_{Al}$ to vary within the ranges 
$-2 \le  \xi_{Al} \le 2$ and $-10 \le  \xi_{Vl} \le 10$. Higher values 
might be also taken into account; the reason why we chose the previous 
ranges was that, to our knowledge, they already include all the most 
known models. 

The results of our calculation are shown in figure 5 \cite{MNPRV}. 
One sees that 
the characteristic features of a general $Z'$ effect are the fact that 
the shifts in the leptonic cross section are essentially negative. This 
can be qualitatively  predicted from the Born approximation formula   
eq.~(\ref{rese4}) because the dominant photon exchange contribution to 
$\sigma_l$ is clearly negative since 
${\widetilde {\Delta}}^{(Z')} \alpha (q^2)$ is negative. Away from 
$\xi_{Al} \simeq 0$ the forward-backward asymmetry will be also negative, 
as easily inferred from eq.~(\ref{rese5}).

\begin{figure}
\vskip -3.5cm

\hbox to\hsize{\epsfxsize=0.50\hsize
\epsffile[0 0 612 792]{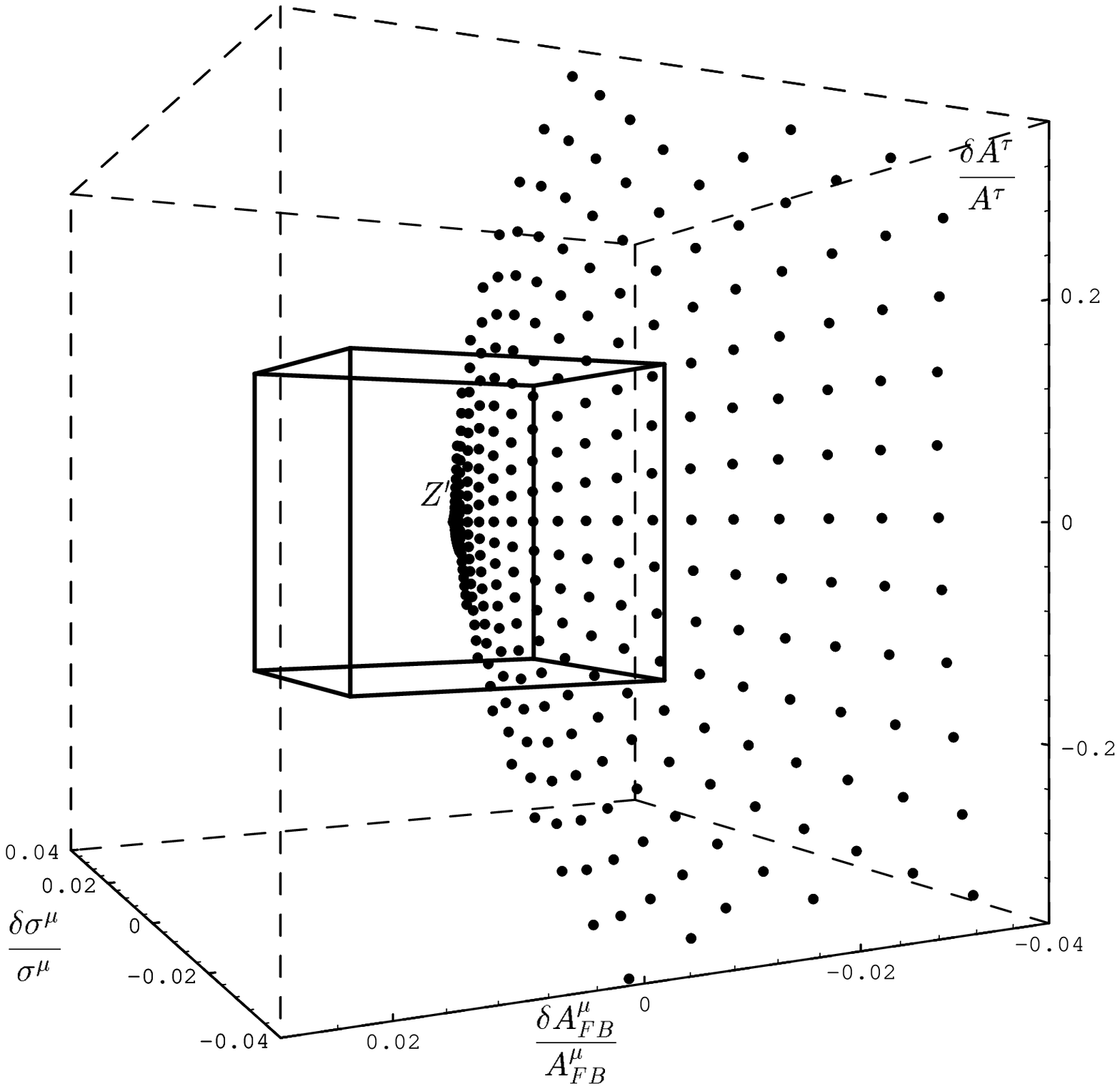}
\hfill\epsfxsize=0.50\hsize
\epsffile[0 0 612 792]{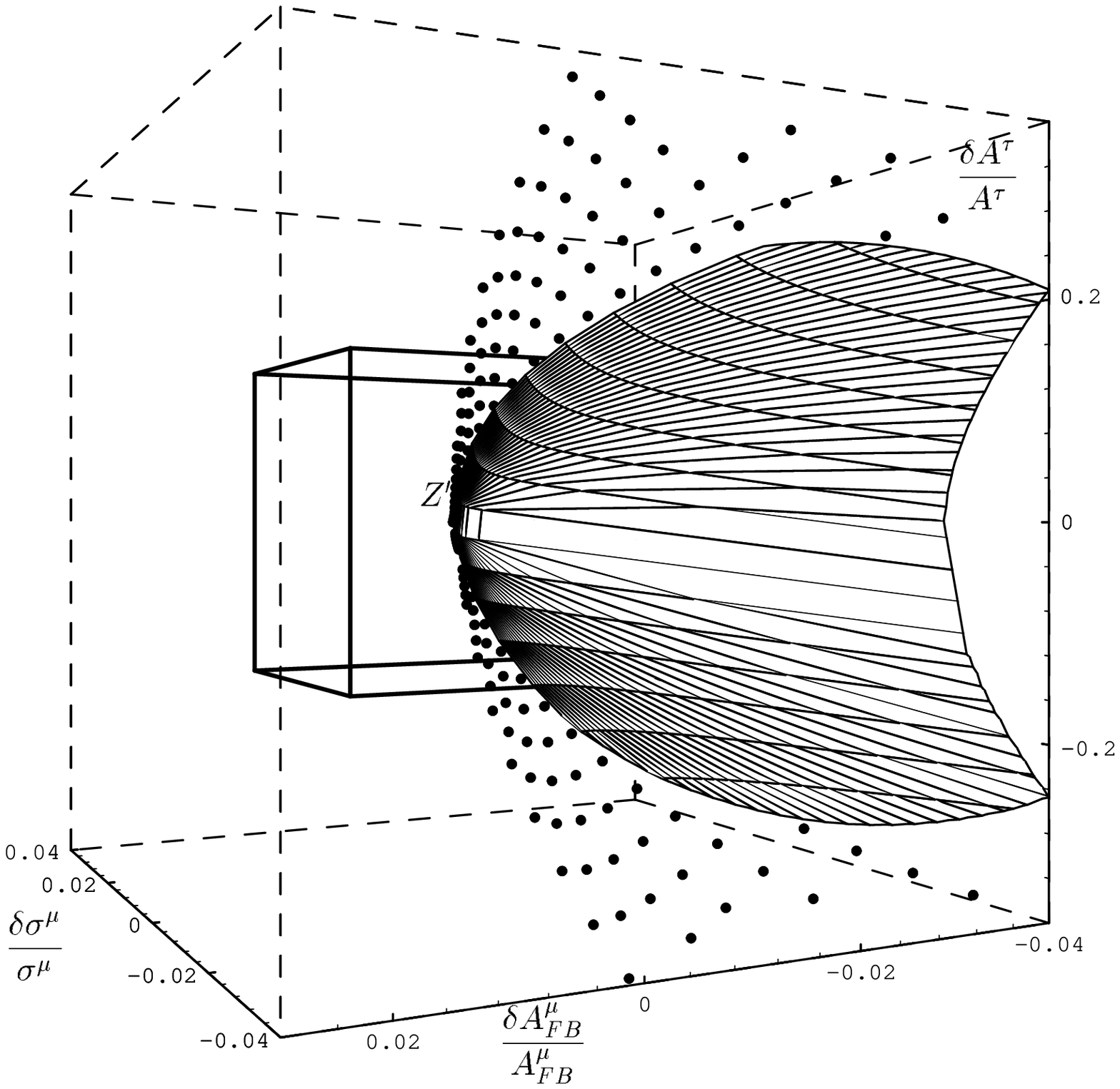}}

\vskip -1cm

\noindent {\bf Fig.~5} {\eight ${{\delta A^\tau} \over {A^\tau}} $ versus 
${{\delta \sigma^\mu} \over {\sigma^\mu}} $  and 
${{\delta A^\mu_{FB}} \over {A^\mu_{FB}}} $. 
The central ''dead'' area
where a signal would not be distinguishable corresponds to an assumed
(relative) experimental error of 1.5\% for $\sigma_\mu$ and to 
1\% (absolute)
errors on the two asymmetries. The region that remains outside 
the dead area
represents the $Z'$ reservation at LEP2, to which the effect of 
the most
general $Z'$ must belong.}\\
\noindent {\bf Fig.~6} {\eight The same as Fig.~5, comparing the realistic
results  obtained via
Monte Carlo simulation with the approximate ones according to 
Born approximation.}

\vskip -2cm
{
\def\epsfsize#1#2{0.41#1}

\hskip 1,8cm\epsfbox{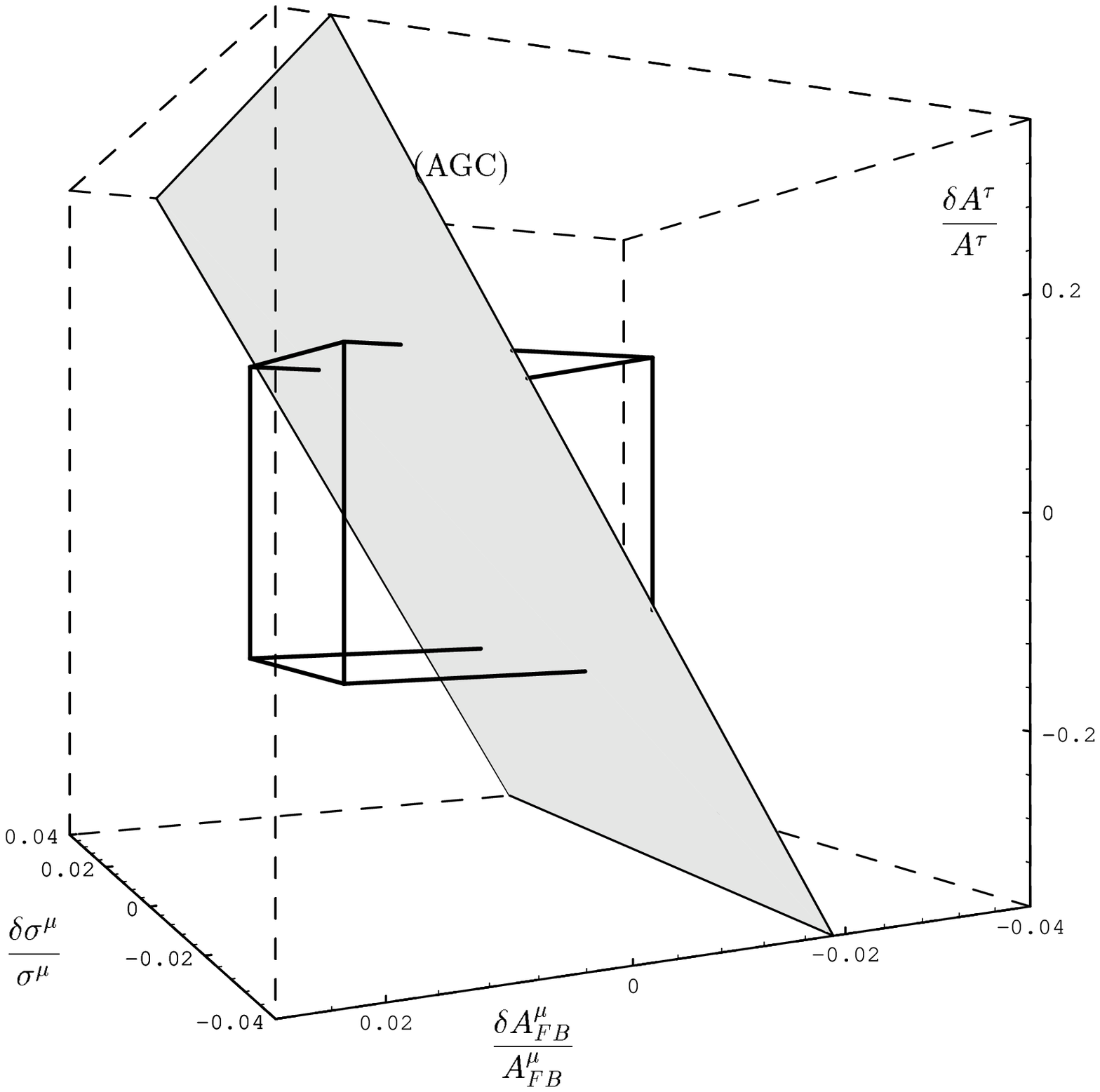}
}

\vskip -2cm

\noindent {\bf Fig.~7} {\eight The region 
corresponding to Anomalous Gauge Couplings according 
to a Born approximation.}


\end{figure}

\newpage

One might be interested in knowing how different the realistic figure 
5 is from the approximate Born one, corresponding to the simplest 
version given in eq.~(\ref{rese7}). This can be seen in figure 6 where 
we have drawn the allowed regions, the points corresponding to the 
realistic situation, already shown in figure 5. The region inside the 
parallelepiped , where a signal would not be detectable, corresponds to 
an assumed relative experimental error of $1.7\%$ for $\sigma_l$ and to 
an absolute error of $1\%$ for $A^l_{FB}$. For the $\tau$ asymmetry an 
absolute error of $2\%$ has been assumed, that is extremely optimistic. 
The domain that remain 
outside this area represents the $Z'$ reservation at LEP2, to which the 
effect of the most general $Z'$ must belong. One sees that the simplest 
Born calculation is, qualitatively, a reasonable approximation to a 
realistic estimate, which could be very useful if one first wanted to 
look for sizeable effects. 

The next relevant question that should be now answered is whether the 
correspondence between $Z'$ and reservation is of the one to one type, 
which would lead to a unique identification of the effect. We have 
tried to answer this question for one specific and relevant case, that 
of virtual effects produced by anomalous gauge couplings. In particular,
we have considered the case of the most  general dimension 6 $SU(2) 
\otimes U(1)$
invariant effective  lagrangian recently proposed \cite{Hagi}. This has
been fully 
discussed in a separate paper \cite{cv3}, where the previously 
mentioned "Z-peak substracted" approach has been used. The resulting 
AGC reservation in the ($\sigma_l$, $A^l_{FB}$, $A_{\tau}$) has been 
calculated for simplicity in the Born approximation, as suggested by 
the previous remarks. This AGC area is plotted in figure 7. As one sees, the 
two domains do not overlap in the meaningful region. Although we cannot 
prove this
property in general, we can at least conclude that, should  a clear virtual
effect
show up at LEP2, it would be possible to decide  unambiguously to which
among two
well known proposed models it does  belong.
  
The results that we have shown so far have been obtained by exploiting the 
information provided by the final fermionic channel. We shall devote the next
section 4 to a brief  discussion of the WW channel at LEP2.

\section{Search for effects in the WW channel.}
The virtual effects of a $Z'$ in W pair production from $e^+e^-$ 
annihilation can be described, at tree level, by adding to the photon, 
 SM Z and neutrino exchanges the diagram with an additional $Z'$ boson  
exchange. The overall effect in the scattering amplitude reads:

\be
A_{lW}^{(0)}(q^2) = A_{lW}^{(0) (\gamma, Z,\nu)}(q^2) + A_{lW}^{(0) Z'}
(q^2)
\label{WW1}
\ee
where we assume universal couplings.
Separate expressions can be easily derived for eq.~(\ref{WW1}). We 
shall only give here the relevant $Z'$ contribution: 
\be 
 A_{lW}^{(0) Z'}(q^2) = \frac{i}{q^2-M^2_{Z'}}  \frac{g_0}{2c_0} 
{\bar v}_l \gamma_{\mu} ({g'}_{Vl}^{(0)}-\gamma^5 {g'}_{Al}^{(0)}) u_l
e_0 g_{ZWW} P^{\alpha \beta \mu} \epsilon^{\star}_{\alpha}(p_1)
\epsilon^{\star}_{\beta}(p_2)
\label{WW2}
\ee
where:
\be 
 P_{\alpha \beta \mu} = g_{\mu \beta}(p_1 + 2 p_2)_{\alpha} + 
 g_{\beta \alpha}(p_1 - p_2)_{\mu} - g_{\mu \alpha}(2p_1 + p_2)_{\beta} 
\ee 
and $p_{1,2}$ are the four momenta of the outgoing W bosons. In the 
expression eq.~(\ref{WW2}) we have assumed that the $Z'WW$ vertex has 
the usual Yang-Mills form. We do not consider here the possibility of 
anomalous magnetic or quadrupole type of couplings. 
 An analysis with anomalous $ZWW$ and $Z'WW$ couplings is possible 
along the lines of \cite{CV4} but is beyond the scope of this report. 
Our analysis will be nevertheless rather general as the trilinear 
$Z'WW$ coupling will be treated as a free parameter, not necessarily 
proportional to the $Z-Z'$ mixing angle as for example it would appear 
in a "conventional" $E_6$ picture.

For the purposes of this working group, it will be particularly 
convenient to describe the virtual $Z'$ effect as an "effective" 
modification of $Z$ and $\gamma$ couplings to fermions and W pairs. As 
one can easily derive, this corresponds to the use of the following modified 
trilinear couplings that fully 
describe the effect in the final  process  $e^+e^- \rightarrow W^+W^-$:
\be
g_{\gamma WW}^{\star} = g_{\gamma WW} +  g_{Z' WW}
\frac{q^2}{M^2_{Z'}-q^2} g_{Vl} (\xi_{Vl} - \xi_{Al})
\label{WW3}
\ee 

\be
g_{ZWW}^{\star} = g_{ZWW} - g_{Z' WW} 
\frac{q^2- M^2_Z}{M^2_{Z'}-q^2} \xi_{Al}
\label{WW4}
\ee 

In the previous equations, the same definitions as in eq.~(\ref{res3}) 
and in eq.~(\ref{res4}) have been used.  In the following we shall use 
the results obtained on $\xi_{Vl}$ and $\xi_{Al}$ in the previous section. 
Our normalisation for  trilinear couplings is such that: 
$ g_{\gamma WW} = 1 $ and
$g_{ZWW} = \frac{c_0}{s_0}$.
 
Adopting the notations that are available in the recent literature 
\cite{KNEUR}, we find for the $Z'$ effect:

\be
 {\delta}_{\gamma}^{(Z')} = g_{\gamma WW}^{\star} - 1 =
   g_{Z' WW}
\frac{q^2}{M^2_{Z'}-q^2} g_{Vl} (\xi_{Vl} - \xi_{Al})
\label{WW5}
\ee

\be
 {\delta}_Z^{(Z')} = g_{ZWW}^{\star} - \cot {\Theta}_W =
 - g_{Z' WW} 
\frac{q^2- M^2_Z}{M^2_{Z'}-q^2} \xi_{Al}
\label{WW6}
\ee 

>From eq.~(\ref{WW5}) and eq.~(\ref{WW6}) one can derive the following 
constraint:

\be
 \frac{{\delta}_{\gamma}^{(Z')}}{{\delta}_Z^{(Z')}} = 
 \frac{- q^2}{q^2-M^2_Z} 
\left( \frac{ \xi_{Vl} - \xi_{Al}}{\xi_{Al}} \right) g_{Vl}  
\label{WW7}
\ee 
We then notice that the virtual effect of a general $Z'$ in the WW 
channel is, at first sight, quite similar to that of a possible model 
with anomalous gauge couplings, that would also produce shifts 
${\delta}_{\gamma}$, ${\delta}_Z$ both in the $\gamma WW$ and in the 
$ZWW$ couplings. But the $Z'$ shifts satisfy in fact the constraint 
given in eq.~(\ref{WW7}), that corresponds to a certain line in the 
(${\delta}_{\gamma}, {\delta}_Z$)  plane whose angular coefficient is 
fixed by the model i.e. by the values of  $\xi_{Al}$ and  
$(\xi_{Vl} - \xi_{Al})$.
 
We shall now introduce the following ansatz concerning the theoretical 
expression of $g_{Z' WW}$, that we shall write as: 

\be
 g_{Z' WW} = \left( c \frac{M^2_Z}{M^2_{Z'}} \right) \cot {\Theta}_W
\label{WW8}
\ee   

The constant c would be of the order of one for the "conventional" 
models where the $Z'$ couples to W only via the $Z-Z'$ mixing (
essentially contained in the bracket of eq.~(\ref{WW8})). But for a 
general model, c could be larger, as one can see for some special cases 
of composite models\cite{Y} or when the $Z'$ originates from a strong coupling 
regime\cite{BESS}.
In fact, a stringent bound on c comes from the request that the partial 
$Z'$ width into WW has to be "small" compared to the $Z'$ mass. 
Imposing the reasonable limit: 

\be
 \Gamma_{Z'WW} \leq \frac{1}{10}{M_{Z'}}
\label{WW9}
\ee 

leads to the condition: 

\be
 c \leq 10
\label{WW10}
\ee
that we consider a rather "extreme" choice.
 
We shall now discuss the observability limits on ${\delta}_{\gamma}$ 
and ${\delta}_Z$. According to \cite{KNEUR}, six equidistant bins in 
the cosine of the production angle are chosen for the generation of 
data, such that each bin contains a reasonable number of events (
$\ge 4$). A binned maximum likelihood method has been used. The result
for one parameter fit ${\delta}_Z$ is:$-0.2 \le  \delta_Z \le 0.25$ for 
the configuration $\sqrt q^2=175$ GeV and $\int L dt=500 pb^{-1}$ and similarly 
for ${\delta}_{\gamma}$. We have then considered a number of possible 
illustrative situations, as extensively discussed in \cite{CRV} and 
found that even in correspondance to the available present 
 CDF limits and for the optimistic choice 
$c=10$, one would get an effect of about $1 \%$, i.e. well below the 
expected LEP2 observability limit.

In conclusion, a $Z'$ of even pathologically small mass, for extreme 
values of its assumed couplings, would be unable to produce observable 
effects in the WW channel at LEP2. Therefore in the derivation of 
bounds or searches for visible effects, the final fermionic channel 
provides all the relevant information.   

\section{Concluding remarks.}

We have tried in this report to be as concise and essential as possible, 
partially owing to the lack of space. In this spirit, we feel that a proper 
conclusion to our work might be that of stressing that LEP2, under 
realistic experimental conditions and in a rather near future, will be able to
 perform a clean and competitive, in some respects quite unique, search 
for effects of a $Z'$ whose mass is not above the TeV boundary. For
$M_{Z'}$ values  beyond this limit, only more energetic machines will be
able to continue this task.

\section*{Acknowledgements.}
This work has been partially supported by EC contract CHRX-CT94-0579.

\end{document}